\documentclass[a4paper,10pt]{article}

\usepackage{times}
\usepackage{amsmath}
\usepackage{amssymb}
\usepackage{bm}
\usepackage{rotating}
\usepackage{float}
\usepackage{color}
\usepackage{multirow}

\newcommand{\nothing}[1]{}

\newcommand{\beq}{\begin{equation}}
\newcommand{\eeq}{\end{equation}}
\newcommand{\bed}{\begin{displaymath}}
\newcommand{\eed}{\end{displaymath}}

\newlength{\defbaselineskip}
\newcommand{\setlinespacing}[1]%
           {\setlength{\baselineskip}{#1 \defbaselineskip}}

\usepackage{amssymb}
\usepackage[ansinew]{inputenc}
\usepackage[english]{babel}
\usepackage{amsfonts}
\usepackage{latexsym}
\usepackage{array}
\usepackage{amssymb}
\usepackage{epsfig}
\usepackage{color}
\usepackage{graphicx}

\usepackage{times}
\usepackage{amsmath}
\usepackage{amssymb}
\usepackage{bm}
\usepackage{rotating}
\usepackage{float}
\usepackage{color}

\def \bX {{\mathbf X}}
\def \bD {{\mathbf D}}
\def \bB {{\mathbf B}}
\def \bW {{\mathbf W}}
\def \bV {{\mathbf V}}
\def \bx {{\mathbf x}}

\def \bt {{\mathbf t}}
\def \bzero {{\mathbf 0}}
\def \bpsi {\mbox{\boldmath $\psi$}}
\def \bphi {\mbox{\boldmath $\phi$}}

\def \bbeta {\mbox{\boldmath $\beta$}}
\def \bEta {\mbox{\boldmath $\eta$}}

\def \bmu {\mbox{\boldmath $\mu$}}

\begin{document}

\begin{center}

\textbf{\Large Disease Mapping via Negative Binomial Regression M-quantiles}

\vspace{1.5cm}

{\large Ray Chambers$^\dag$, Emanuela Dreassi$^\ddag$, Nicola Salvati$^\S$}

\vspace{0.5cm}

$\dag$ National Institute for Applied Statistics Research Australia, \\University of Wollongong, Australia\\
$\ddag$ Dipartimento di Statistica, Informatica, Applicazioni (DiSIA), \\Universit\`a degli Studi di Firenze, Firenze, Italy\\
$\S$ Dipartimento di Economia e Management, \\Universit\`a di Pisa, Pisa, Italy\\

\end{center}

\begin{abstract}
We introduce a semi-parametric approach to ecological regression for disease mapping,
based on modelling the regression M-quantiles of a Negative Binomial variable.
The proposed method is robust to
outliers in the model covariates, including those due to measurement error,
and can account for both spatial heterogeneity and spatial clustering.
A simulation experiment based on the well-known Scottish lip cancer data set
is used to compare the M-quantile modelling approach and
a random effects modelling approach for disease mapping. This suggests that
the M-quantile approach leads to predicted relative risks with smaller root mean
square error than standard disease mapping methods.
The paper concludes with an illustrative application of the M-quantile approach,
mapping low birth weight incidence data for English Local Authority
Districts for the years 2005-2010.
\end{abstract}

\vspace{0.5cm}

\textbf{keywords:} Ecological regression; Overdispersed count data; Robust models; Spatial correlation

\section{Introduction}
\label{intro}

Disease mapping involves the analysis of disease incidence or
mortality data for a specified geographical region that has been subdivided into small areas. These data are typically area level counts, and are usually combined with data on area level covariates that could be considered as risk factors when assessing how the associated relative risks vary from area to area.

Ecological regression is the analysis of the association between risk factors and disease incidence for these areas, while disease mapping is the estimation of their disease risk, based on ecological regression models. The area level counts used for this purpose typically exhibit overdispersion, and an Empirical Bayes approach (referred to as EB below) that uses a Poisson-Gamma model for relative risks was proposed by Clayton and Kaldor \cite{clakal}. Subsequently, a Hierarchical Bayes generalization of this approach that allows for a spatial structure was developed by Besag \emph{et al.} \cite{bym} (hereafter BYM). Ecological disease mapping
typically relies on regression models that use covariates to explain risk variation between areas and
random effects to allow for this overdispersion. These models depend on
distributional assumptions and require a formal specification of the
random part of the model. Furthermore, applications involving spatially heterogeneous
data require predictors that are more flexible than the usual linear
predictor (see, for example, space varying coefficients models:
\cite{assuncao} and \cite{laws}), while standard ecological regression models do not easily
allow for outlier-robust inference, e.g. when outliers are due to the presence of area level covariates with measurement error (e.g. \cite{berna97}, \cite{xia98}, \cite{mcnab09} and \cite{mcnab10}).

Ecological regression for disease mapping can be regarded as a special case of
small area estimation \cite[Chapter 9]{rao}.
In particular, the EB predictor of relative risk for an area belongs to the family of small area estimators defined by generalized linear mixed models. This family includes a wide variety of different models, ranging from models for binary and count data to models for a continuous response, e.g.
linear mixed models with Gaussian residuals. In the latter case, EB and Empirical Best Linear Unbiased Predictor (EBLUP) estimators
coincide \cite[Chapter 9]{rao}. For the case of a continuous response, Chambers and Tzavidis
\cite{chatza} proposed an approach to small area estimation based on linear regression M-quantiles. This
approach involves weaker parametric assumptions than the linear mixed model, and is robust to outliers in the response because of its use of M-estimation.

In this paper, we define regression M-quantiles for count data that can be characterised as Negative Binomial, focussing on applications to ecological regression for disease mapping. This is referred to as the NBMQ approach below. Furthermore, since the data that are used in such applications typically exhibit spatial clustering, we extend the method to allow for the presence of this clustering, referring to it as NBMQsp below.
As with other applications of M-quantile modelling for grouped data,
the NBMQ approach does not use random effects to characterise groups, which in this case
correspond to areas. Instead, between area variation in the response is characterised by variation
in area-specific values of quantile-like coefficients. Furthermore, since this approach is based on an
outlier-robust approach to fitting generalised linear models, it leads to outlier-robust inference
when area level covariates are measured with error.

We compare the NBMQ approach with the EB and BYM approaches
using a simulation experiment based on the well known Scottish lip cancer data. The results suggest that
the new approach generates estimates of disease prevalence with smaller root mean square error
than those generated using these standard mixed model based approaches to disease mapping. We also illustrate application of NBMQ for disease mapping by comparing it with EB and BYM when mapping low birth weight incidence rates for English local authorities.

The paper is organized as follows. In Section \ref{over}, the
Negative Binomial model for overdispersed count data and
disease mapping is reviewed. In Section \ref{rob}, the robust Negative
Binomial model, which extends the class of models introduced by Cantoni and Ronchetti
\cite{cantoronche}, is described. In Section \ref{mq}, the
NBMQ model for overdispersed count data is
introduced and applied to disease mapping. This section also contains a description of
the NBMQsp approach, which extends the disease-mapping application of NBMQ to data that
exhibit spatial clustering, and a description of a semiparametric bootstrap method for estimating
the MSE of both NBMQ predictors. Results from a simulation study that compares NBMQ, EB and BYM
with respect to their bias and root mean squared error are discussed in Section \ref{sim}.
In Section \ref{exe}, the method is illustrated through an example: low birth weight
incidence data for $326$ local authority districts of England, during the period 2005-2010.
Finally, in Section \ref{con} we draw some conclusions about the usefulness of the NBMQ approach, and identify areas for further research.

\section{Overdispersed count data}
\label{over} The Poisson distribution is useful for
modelling the mean behaviour of count data but can underestimate variability when these data are overdispersed.
There are essentially three ways for dealing with this situation. One
is to use the Poisson maximum likelihood
estimating function for the mean, but to then base inference on the heterogeneity
robust sandwich covariance matrix estimator. The second is to use a Quasi-Poisson model (see \cite{cametrive}).
The third is to model the overdispersed count data directly using a Negative Binomial
model, i.e. as a Gamma mixture of Poisson
distributions. We focus on this third approach in this paper.

Let $Y \sim \mbox{Poisson}(\lambda)$ with $\lambda \sim
\mbox{Gamma}(\theta, \alpha)$. The distribution generated by this compound process is called the Negative Binomial (NB) and has density

\[
p(y; \alpha, \theta)=\left(\begin{array}{c}
                        y+\theta-1 \\
                         \theta-1
                      \end{array}\right)
                      \left( \frac{\alpha}{1+\alpha}\right)^{\theta}
                      \left( \frac{1}{1+\alpha}\right)^y
\]
where $y=0, 1, 2, \ldots$ can be characterized as the number of failures before $\theta$
successes, with success probability $p=\alpha/(1+\alpha)$. The mean and variance of this distribution is $\mbox{E}[Y]=\theta / \alpha$ and $\mbox{Var}[Y]=\theta /
\alpha+\theta/\alpha^2$. We reparameterize, setting
$\mu=\theta/\alpha$, to obtain
\[
p(y; \mu, \theta) =\frac{\Gamma(y+\theta)}{\Gamma(\theta) y!} \left(\frac{\theta}{\mu+\theta}\right)^\theta \left(\frac{\mu}{\mu+\theta}\right)^{y}
\]
where now $E[Y]=\mu$ and
$\mbox{Var}[Y]=\mu+\frac{\mu^2}{\theta}$. Since the overdispersion (relative to the Poisson model) in this distribution is a quadratic function of the mean, it is
referred to as the NEGBIN2 or NB2 model in \cite{cametrive}. The value
$1/\theta$ is directly related to the amount of overdispersion in
the data: smaller values of $\theta$ suggest increasing
amounts of overdispersion.

In the context of ecological regression, $Y$ is a count and $\bx$ is a $p \times 1$ vector of explanatory variables (which is assumed to include the
constant term). The regression of $Y_i$ on $\bx_i$ is modelled as $\mu(\bx_i)= \exp \eta_i =
\exp(\bx^T_i \bbeta)$, where $\bbeta$ is a vector of
$p$ regression parameters. Given $n$ observations this regression model can be written as $\log(\bmu)=\bEta=\bX \bbeta$. Since the NB distribution is a member of
the exponential family for fixed $\theta$, this model is a special case of the
Generalized Linear Model (GLM), with the $\log(\cdot)$ link function. In line with standard practice (\cite{mccullagh},
\cite{breslow}, \cite{lawless}), GLM methodology can be used to estimate $\bbeta$, by replacing $\theta$ with a suitable estimate
$\hat{\theta}$ (obtained using the method of moments, for example) and by
iterating estimation of $\bbeta$ given $\hat{\theta}$.

Log-linear ecological regression models for count data are the basic building blocks for estimating relative risk of disease (including mortality) from incidence data. In many applications these data are
available at an aggregated geographic level, e.g. corresponding to a defined area on a map. In the next section we review these
`standard' disease mapping methods, with the aim of using them as benchmarks for the NBMQ and NBMQsp methods that we introduce later.

\subsection{Models for disease mapping}

Consider a region partitioned into $n$ distinct areas, and let $y_i$ denote a count
associated with area $i=1, \ldots, n$, e.g. the number of recorded cases of a disease or the number of deaths. Each $y_i$ is
assumed to be an independent realization of a random variable $Y_i \sim \mbox{Poisson}(\mu_i)$, where $\mu_i=t_i \lambda_i$. Here $t_i$ is a baseline expected count in area $i$ and $\lambda_i$ is the relative risk. The MLE for $\lambda_i$ is $y_i /t_i$. However, since such
data are characteristically overdispersed, James-Stein type
estimators are preferred (see \cite{efron}). Following Clayton and Kaldor \cite{clakal} the $\lambda_i$ are
assumed to be independently and identically distributed as
Gamma$(\theta, \alpha)$. The resulting compound model is a NB
model with mean $\theta (t_i / \alpha)$ and variance $\theta (t_i/
\alpha) + \theta (t_i/\alpha)^2$. Conditionally on the values of the other model
parameters and the data, each $\lambda_i$ then has a posterior Gamma distribution
with mean $\mbox{E}[\lambda_i \mid y_i, \theta,
\alpha]=(y_i+\theta)/(t_i+\alpha)$. The empirical Bayes (EB)
estimator of $\lambda_i$ is the corresponding plug-in estimator of this parameter, defined by replacing $\alpha$ and $\theta$ in this posterior mean by
suitable estimates (e.g. their MLEs). Clearly, we can extend this to an ecological regression model by making $\lambda_i$, and hence $\mu_i$, a function of a set of
covariates.

The EB method has been extended to a Hierarchical Bayes (HB) approach by Besag \emph{et al.} \cite{bym}.
Their standard model is of the form
\begin{equation}
\label{HBstandard}
\log(\lambda_i) = \beta_0+ \sum_{j=1}^{p-1} \beta_j
x_{ij} + u_i + v_i
\end{equation}
where $\beta_0$ represents an intercept, such as an overall risk
level; $\beta_1, \ldots, \beta_{p-1}$ is a set of regression
coefficients; $u_{i}$ is a spatially correlated random effect (the
\emph{clustering} effect), and $v_{i}$ is a spatially uncorrelated
random effect (the \emph{heterogeneity} effect). Prior distributions
for the model parameters are typically specified as follows: the
intercept $\beta_0$ is assumed to have a uniform prior
distribution; the coefficients $\beta_j$ are assumed to have a
normal prior distribution with zero mean and small precision; the
heterogeneity effects $v_{i}$ are assumed to be independently
distributed as normal with mean $0$ and variance $\tau_{v}^{-1}$;
and the clustering effects $u_{i}$ are assumed to be realisations of
a Gaussian Markov Random Field (GMRF), which is modelled by
conditioning on the values of spatially neighbouring clustering
effects in the sense that $u_{l \sim i }$ is assumed to follow a
Normal $(\bar{u_{i}}, (\tau_{u} m_i)^{-1})$ distribution, where
$\bar{u_i}=\sum_{l \sim i} \frac{u_{l}}{m_i}$. Here $l \sim i$ denotes areas that are adjacent to area $i$
(i.e. areas that share a boundary with area $i$) and $m_i$ is the
number of areas that are adjacent to area $i$. The parameters
$\tau_{v}$ and $\tau_{u}$ are typically assumed to have gamma
priors, see \cite{kensa} for further details. The marginal
posterior distributions of the parameters of interest are then
approximated by Monte Carlo Markov Chain methods. We refer to the HB estimates based on fitting this model by BYMsp in what follows. Note that (\ref{HBstandard})
can also be fitted without a spatial clustering effect (i.e. just with the heterogeneity effect $v_{i}$). We use BYM to refer to HB estimates based on such a fit in what follows.

\section{Robust estimation for the Negative Binomial model}
\label{rob}

Cantoni and Ronchetti \cite{cantoronche} propose an approach to robust inference for generalized linear models based on quasi-likelihood.
In particular, they consider a general class of M-estimators of Mallows's type, where the influence of deviations on $y$ and on $\bX$ are bounded separately. Their robust version of the estimating equations for the parameter $\bbeta$ of the GLM is of the form
\begin{equation} \label{CReq}
n^{-1}\sum_{i=1}^{n} \bphi(y_i, \mu_i) =\bzero
\end{equation}
where $\bphi(y_i, \mu_i)=v(y_i,
\mu_i)w(\bx_i)\mu_i^{\prime}-a(\bbeta)$, $E[Y_i]=\mu_i$,
$V[Y_i]=V(\mu_i)$, $\mu_i=\mu_i(\bbeta)=g^{-1}(\bx_i^{T}\bbeta)$,
$\mu_i^{\prime}$ is its derivative and
$a(\bbeta)=\frac{1}{n}\sum_{i=1}^n
E[v(y_i,\mu_i)]w(\bx_i)\mu_i^{\prime}$ ensures the Fisher
consistency of the estimator. The function $v(y, \mu)$ is a bounded function of model residuals that
controls the influence of errors in $y$-space, whereas the weights
$w(\bx)$ are used to downweight leverage points. When
$w(\bx_i)=1~\forall~i$ Cantoni and Ronchetti \cite{cantoronche} call the estimator defined by the solution to (\ref{CReq}) the
Huber quasi-likelihood estimator, using it to obtain robust
estimates for parameters of Binomial and Poisson models in the case where $v(y, \mu)$ is defined by Pearson
residuals and the Huber influence function. Note that the solution to (\ref{CReq}) can be obtained numerically by a
Fisher scoring procedure.

We extend this approach to robust fitting of the mean parameterized NB model, via the estimating equations
\begin{equation} \label{CRNB2eq}
\Psi(\bbeta):=n^{-1} \sum_{i=1}^{n}\bpsi(y_i, \mu_i) =\bzero
\end{equation}
where $\bpsi(y_i, \mu_i)=\Big \{\psi(r_i) w(\bx_i)
\frac{1}{V^{1/2}(\mu_i)}\mu_i^{\prime}-a(\bbeta) \Big \}$,
$r_i=\frac{y_i-\mu_i}{V^{1/2}(\mu_i)}$ are the Pearson residuals,
$\psi(\cdot)$ is the Huber Proposal 2 influence function, $\psi(r)= r
\, I(-c < r < c) + c \, \mbox{sgn}(r) \, I(|r| \geq c)$, $c$ is the tuning constant,
$\mu_i=t_i \exp{(\bx_i^{T}\bbeta)}$, $t_i$ is the offset term,
$\mu_i^{\prime}=\mu_i \bx_i^{T}$,
$V(\mu_i)=\mu_i+\frac{\mu_i^2}{\theta}$ and $\theta>0$ is a shape
parameter. The correction term $a(\bbeta)=1/n \sum_{i=1}^n
E[\psi(r_i)] V^{-1/2} (\mu_i) w(\bx_i) \mu_i^{\prime}$ can be computed
explicitly for the NB model, as shown in Appendix.
In order to ensure that the solution to (\ref{CRNB2eq}) is robust, the parameter $\theta$ is estimated using a robust method. We propose the use of the robust scale estimator
\cite{huber81} defined by
\begin{equation}\label{NB2theta}
n^{-1}\sum_{i=1}^{n}\left\{ \psi^2(r_i)- E \left[ \psi^2 \left(
\frac{ Y_i-\mu_i}{V^{1/2}(\mu_i)}\right) \right]\right\}=\bzero,
\end{equation}
where $E \left[ \psi^2 \left( \frac{
Y_i-\mu_i}{V^{1/2}(\mu_i)}\right) \right]$ is a constant that ensures that the solution to (\ref{NB2theta}) is Fisher consistent (see the Appendix for its computation) and $\psi$ is chosen as in (\ref{CRNB2eq}). The equations (\ref{CRNB2eq}) and (\ref{NB2theta}) can be solved by iterating between a solution to (\ref{CRNB2eq}) given $\theta$ and a solution to (\ref{NB2theta}) given $\bbeta$.

Following Cantoni and Ronchetti \cite{cantoronche} we can write down a sandwich-type approximation to the variance of the solution to (\ref{CRNB2eq}) as
\begin{equation}\label{sandwich}
\mbox{Var}(\hat{\bbeta})\approx\bW^{-1} \bV (\bW^{T})^{-1}.
\end{equation}
Here
\[
\bV=\frac{1}{n}\bX^{T}\bD\bX-a(\bbeta)a(\bbeta)^{T},
\]
where $\bD$ is a diagonal matrix with elements $d_i=E[\psi^2(r_i)]w^{2}(\bx_i)\frac{1}{V(\mu_i)} \Big( \frac{\partial \mu_i}{\partial \eta_i} \Big)^2$ and
\[
\bW=\frac{1}{n}\bX^{T}\bB\bX,
\]
where $\bB$ is a diagonal matrix with elements $b_i=E[\psi(r_i)\frac{\partial \log(h(y_i; \theta, \mu_i))}{\partial \mu_i}]\frac{1}{V^{1/2}(\mu_i)}w(\bx_i)\Big( \frac{\partial \mu_i}{\partial \eta_i} \Big)^2$, with $h(\cdot)$ the conditional density of $y_i|\bx_i$ and
$\frac{\partial \log(h(y_i; \theta, \mu_i))}{\partial \mu_i}=\sum_{i=1}^n \frac{y_i-\mu_i}{V(\mu_i)}$. Computational formulae for the elements of $\bD$ and $\bB$ are set out in the Appendix.
An estimator of the first order approximation (\ref{sandwich}) is then
\begin{equation}\label{sandwich_est}
\widehat{\mbox{Var}}(\hat{\bbeta})=\hat{\bW}^{-1} \hat{\bV}(\hat{\bW}^{T})^{-1}.
\end{equation}

\section{Regression M-quantiles for Negative Binomial data}
\label{mq}

The M-quantiles of a random variable $Y$ with continuous distribution function $F(\cdot)$ are a `quantile like' characterisation of $F$ and were introduced in \cite{breckcha}, who noted that the relationship between an M-estimate of the location of $F$ and its corresponding sample M-quantiles is the same as that between its sample median and corresponding sample quantiles. M-quantile regression is a generalization of regression to the M-quantiles of the conditional distribution of $Y$ given a vector $\bx$ of covariates. In particular, the regression M-quantile of order $q$ for this distribution, $q\in (0,1)$, is defined as the solution $Q_q (\bx; \psi)$ to

\begin{equation}
\label{Mquantile}
E \left[ \psi_q \left(\frac{Y-Q_q (\bx; \psi)}{\sigma_q} \right) \right] =0,
\end{equation}

\noindent where the expectation is conditional on $\bx$, $\psi_q (r)=2 \psi ( r /
\sigma_q) \left[ q \, I(r > 0) + (1-q) I (r \leq 0 ) \right]$, $\sigma_q$ is the scale of the random variable $Y-Q_q (\bx; \psi)$,
and $\psi$ is an appropriately chosen influence function. A linear regression M-quantile of order $q$ satisfies $Q_q (\bx; \psi)=\bx \bbeta_{q}$, where $\bbeta_{q}$ is then the $p \times 1$ vector of regression
coefficients that defines the M-quantile of order $q$ of the conditional distribution of $Y$ given $\bx$. By analogy with standard M-regression, an estimator of
$\bbeta_{q}$ can be obtained as the solution to the set of estimating
equations
\begin{equation}\label{lmqee}
n^{-1}\sum_{i=1}^{n}\psi_q(\frac{r_{iq}}{s})\bx_i=\bzero,
\end{equation}
where $r_{iq}=y_i-\bx_i^{T}\bbeta_q$ and $s$ is a  robust  estimate of scale, e.g. the median absolute deviation estimate $s = \mbox{median} \mid r_{i}^{\mbox{ols}} \mid /0.6745$, where the $r_{i}^{\mbox{ols}}$ are the residuals generated by an OLS fit. It is straightforward to obtain a solution to (\ref{lmqee}) using an iteratively re-weighted least squares algorithm. Furthermore, if $\psi$ is continuous and monotone non-decreasing (e.g. a Huber-type function) then this algorithm is guaranteed to converge to a unique solution \cite{Kokic1997}.

\subsection{M-quantile regression for a count response}
The quantile function of a discrete random variable is not generally a monotone increasing function of $q$, so a unique solution to (\ref{Mquantile}) for distinct values of $q$ does not exist if $Y$ is a count and $\psi(r)=sgn(r)$, i.e. the influence function corresponding to the median. However, this is not the case if $\psi$ is a continuous monotone non-decreasing function, in which case a unique solution always exists provided the expectation exists. This allows the concept of regression M-quantiles to be extended to count data in a straightforward way. In the case of NB data, an appealing model for $Q_q (\bx; \psi)$ is then
\begin{equation}\label{NBMQModel}
Q_q (\bx; \psi)=\bt \exp (\bEta_{q}),
\end{equation}
where $\bEta_{q}=\bx ^T\bbeta_{q}$ is the linear predictor and
$\bt$ is a vector of offset terms (expected or baseline cases) that can potentially be included in the model.

In order to estimate $\bbeta_{q}$ we consider the extension of (\ref{CRNB2eq}) to the M-quantile case. In particular, we replace $\mu_i$ there by $Q_q (\bx_i; \psi)$, leading to the estimating equations
\begin{equation} \label{CRNB2MQeq}
\Psi(\bbeta_q):=n^{-1}\sum_{i=1}^{n}\bpsi_q(y_i, Q_q (\bx_i; \psi)) =\bzero,
\end{equation}
where $\bpsi_q(y_i, Q_q (\bx_i; \psi))=\Big[\psi_q(r_{iq}) w(\bx_i)
\frac{Q_q ^{\prime}(\bx_i; \psi)}{V^{1/2}(Q_q(\bx_i; \psi))}
-a(\bbeta_q) \Big]$, $r_{iq}=\frac{y_i-Q_{q} (\bx_i;
\psi)}{V^{1/2}(Q_{q} (\bx_i; \psi))}$, $V(Q_{q} (\bx_i; \psi))=Q_{q}
(\bx_i; \psi)+\frac{Q_{q} (\bx_i; \psi)^2}{\theta_q}$, $\theta_q>0$
is a shape parameter and  $Q_{q}^{\prime} (\bx_i; \psi)=Q_{q} (\bx_i;
\psi)\bx_i$. Furthermore, using the results in the Appendix for
robust NEGBIN2,
\[
\begin{array}{rcl}
  a(\bbeta_q)&=&n^{-1}\sum_{i=1}^{n} w_q(r_{iq}) w(\bx_i)\left \{ -c \, \, P\left(Y_i \leq
j_1\right) + c \, P \left(Y_i \geq j_2+1 \right)  \right. \\
& +& \left. \frac{Q_{q} (\bx_i; \psi)}{V^{1/2}(Q_{q} (\bx_i; \psi))} P( Y_i=j_1) \left( 1+\frac{j_1}{\theta_q}
\right) \right. \\
 &-&  \left.  \frac{Q_{q} (\bx_i; \psi)}{V^{1/2}(Q_{q} (\bx_i; \psi))} P( Y_i=j_2) \left(
1+\frac{j_2}{\theta_q} \right) \right\}V^{-1/2}(Q_{q} (\bx_i; \psi))Q_{q} (\bx_i;
\psi)\bx_i,
\end{array}
\]
where $j_1=\lfloor Q_{q} (\bx_i; \psi)-c V^{1/2}(Q_{q} (\bx_i;
\psi)) \rfloor$, $j_2=\lfloor Q_{q} (\bx_i; \psi)+c V^{1/2}(Q_{q}
(\bx_i; \psi)) \rfloor$ and $w_q(r_{iq})=2 \, [q \, I(r_{iq}>0)+(1-q)
I(r_{iq} \leq 0)]$. As noted earlier, the equations (\ref{CRNB2MQeq}) can be solved using
Fisher scoring, with the parameter
$\theta_q$ estimated analogously to (\ref{NB2theta}) as the solution to
\begin{equation}\label{NB2thetaMq}
n^{-1}\sum_{i=1}^{n}\left\{\psi_q^2(r_{iq})- E \left[ \psi_q^2
\left( \frac{ Y_i-Q_{q} (\bx_i; \psi)}{V^{1/2}(Q_{q} (\bx_i;
\psi))}\right) \right]\right\}=\bzero,
\end{equation}
where $E \left[ \psi_q^2 \left( \frac{ Y_i-Q_{q} (\bx_i;
\psi)}{V^{1/2}(Q_{q} (\bx_i; \psi))}\right) \right]$ is a constant
that ensures Fisher consistency for estimation of $\theta_q$ and
$\psi_q$ was defined following (\ref{CRNB2MQeq}).
Routines in \texttt{R} that solve these estimating equations are
available from the authors, and we refer to the `ensemble' model defined by the solutions to (\ref{CRNB2MQeq}) and (\ref{NB2thetaMq}) for a range of values of $q$ as a NBMQ model below.

We note in passing that all fitted regression M-quantiles are potentially subject to the
phenomenon of quantile crossing. Theoretically, regression M-quantiles are strictly non-decreasing in $q$ at every $\bx$. However, in practice two or more
fitted regression M-quantiles can sometimes `cross' in the sense that this non-decreasing property does not hold at every sample value of $\bx$.
He \cite{he1997} proposed a posteriori adjustments to fitted regression quantiles to
eliminate crossing, and Pratesi \emph{et al.} \cite{pratesi:2009} adapted
this procedure to p-spline regression M-quantiles. Our implementation of regression M-quantiles based on (\ref{NBMQModel}) could use the approach proposed by He \cite{he1997} to define NBMQ curves that do not cross.

\subsection{Using NBMQ models for disease mapping}
HB models like (\ref{HBstandard}) characterise the variability associated with
the conditional distribution of an overdispersed count variable $Y$ given covariates $\bx$ in terms of
latent clustering and heterogeneity effects. However, a NBMQ model can also be used to characterise this overdispersion by associating a unique `M-quantile coefficient'  with
each observed count.

The M-quantile coefficient associated with the observed value $y_i$ of a continuously distributed random variable $Y$ and an associated covariate value $\bx_i$ is the value $q_i$ such that
$\hat{Q}_{q_i} (\bx_i; \psi) = y_i$, see \cite{chatza}. Typically, this equation is solved by fitting the regression M-quantiles on a finite grid $G=\{0<q_1<q_2<\ldots<q_{L-1}<q_L<1\}$ of $L$ values of $q$ and using linear interpolation. Unfortunately, with NB data and $Q_q (\bx_i; \psi)$ defined by
(\ref{NBMQModel}), this definition cannot be used without modification since an observed count $y_i=0$ can never be part of the strictly positive domain of $Q_q (\bx_i; \psi)$. To overcome this problem we use the
following definition:
\[
\hat{Q}_{q_i}
(\bx_i; \psi)= \left \{\begin{array}{ll}k(\bx_i) & y_i=0 \\y_i & y_i=1,2,\dots \end{array} \right.
\]
where $k(\bx)$ denotes an appropriate strictly positive boundary function for the data set. Note that this function cannot be its convex hull, since that will take the value zero where $y=0$. Another possibility is $k(\bx)=\hat{Q}_{q_1} (\bx; \psi)$. However this implies that the M-quantile coefficient for any value $y=0$ is $q_1$, irrespective of the value of $\bx$. One way to tackle this issue is to argue that the observation $y_i=0$ corresponds to a smaller $q$-value then the observation $y_j=0$ when $\hat{Q}_{0.5} (\bx_i; \psi)>\hat{Q}_{0.5} (\bx_j; \psi)$. As a consequence, we put $k(\bx)=\mbox{min} \{ 1-\epsilon, [\hat{Q}_{0.5} (\bx; \psi)]^{-1} \}$, where $\epsilon>0$ is a small positive constant. For $L>1$, this value can be set equal to $-\mbox{median}(\bx_i^{T} \bbeta_{0.5}),~i=1,\ldots,n$, so that approximately half the observations with y = 0 have $q > 0.5$ and the remainder have $q \leq 0.5$. The
M-quantile coefficient associated with $y_i$ and $\bx_i$ is then $q_i$, where
\begin{equation}\label{qvalue}
\hat{Q}_{q_i}
(\bx_i; \psi)= \left \{\begin{array}{ll} \mbox{min} \Big \{1-\epsilon, \frac{1}{t_i \exp(\bx_i^{T}\hat{\bbeta}_{0.5})}    \Big\}& y_i=0 \\y_i & y_i=1,2,\ldots \end{array} \right.
\end{equation}

Focusing on the choice of the grid $G$ used to solve (\ref{qvalue}), we observe that BYM `borrows strength' via specification of the variance parameter $\tau_v$. In effect, as $\tau_v$ goes to infinity one ends up in the limit with GLM behaviour, which, in the NBMQ case, corresponds to a single point grid $G$ with $q_1=0.5$. This implies that one way of capturing area heterogeneity in a NBMQ model is via specification of $G$. If we note that the $q$ values of the empirical sample quantiles of a distribution are defined by the set $G_n=\left\{ \frac{1}{(n+1)},\ldots,\frac{n}{(n+1)}\right\}$ irrespective of the variance of this distribution, then $G_n$ is a robust default definition for the grid $G$.

In environmental and epidemiological applications, observations that are spatially close may be more alike than observations that are further apart. Equivalently, M-quantile coefficients of observations that are spatially close should be similar. One way of incorporating this information is to spatially smooth the grid-based solutions to (\ref{qvalue}). This can be done in a variety of ways. For example, we can average grid-based solutions $q_l$ from adjacent areas using the formula
\begin{equation}\label{sp_qvalue}
q_i^{\mbox{sp}}=\frac{q_i+m_i^{-1}\sum_{l \sim i} q_l }{2}.
\end{equation}
Other kinds of spatial averaging are possible. For example, we can average using weights $w(d_{il})$ whose values depends on the Euclidean distance between the centroids of the areas $i$ and $l$ so that areas close to area $i$ receive more weight than those further away. In this case the spatially averaged M-quantile coefficient $q_i^{\mbox{sp}}$ becomes
\[
q_i^{\mbox{sp}}=\frac{\sum_{l=1}^n q_l \, w(d_{il})}{\sum_{l=1}^n w(d_{il})}.
\]
A simple Gaussian specification for this weighting function is $w(d)=\exp (- d^2/2 b^2)$, where $b>0$ is the bandwidth. In this case the spatial weight $w(d_{il})$ decreases exponentially as the distance $d_{il}$ increases, with the
bandwidth $b$ determining the speed of this decay.

Following Chambers and Tzavidis \cite{chatza} we then note that the M-quantile coefficients defined by a NBMQ model can be used to capture residual between-area
variation by the deviation of the area-specific M-quantile
regression coefficient $\bbeta_{q_i}$ from the `median' M-quantile
coefficient $\bbeta_{0.5}$. In particular, the NBMQ predictor of the count in area $i$ is then
\begin{equation}\label{prediction}
\hat{Q}_{q_i} (\bx_i; \psi)=t_i \exp (\bx_i^T\hat{\bbeta}_{q_i}).
\end{equation}
The spatial version of (\ref{prediction}), referred to as NBMQsp below, is defined by replacing $q_i$ by $q_i^{\mbox{ sp}}$.

Finally, we observe that we can write the NBMQ model in a form that mimics the
HB model (\ref{HBstandard}), via the identity
\begin{equation}\label{pseudo}
Q_q (\bx_i; \psi)=t_i \exp (\bx_i^T \bbeta_{0.5}+\bx_i^T(\bbeta_{q_i}-\bbeta_{0.5})).
\end{equation}
The last term on the right-hand side of (\ref{pseudo}) can be interpreted as a pseudo-random effect for area $i$, allowing estimation of area effects.

\subsection{Mean squared error estimation}
We propose a semiparametric bootstrap-based estimator for estimating the MSE of  (\ref{prediction}). This bootstrap is semiparametric in nature in the sense that area effects are generated using an empirical rather than a parametric distribution. Given the $n$ values of the count variable $y_{i}$ the steps of this bootstrap procedure are summarized as follows:
\begin{itemize}
\item [step 1.] Fit the model (\ref{NBMQModel}) to the data to obtain a predicted value $\hat{Q}_{q_i} (\bx_i; \psi)$, an estimated pseudo-random effect $\hat{u}_i^{\mbox{ NBMQ}}=\bar{\bx}_i^T(\hat{\bbeta}_{q_i}-\hat{\bbeta}_{0.5})$ and an estimate $\hat{\theta}_{q_i}^{\mbox{ NBMQ}}$ of the shape parameter for each area $i$. It is convenient to re-scale the $\hat{u}_i^{\mbox{ NBMQ}}$ so that they have mean exactly equal to zero.
\item [step 2.] Construct the sets $\{ \hat{u}_i^{\mbox{ NBMQ} \ast}; i=1,\dots,n\}$ and $\{ \hat{\theta}_i^{\mbox{ NBMQ} \ast}; i=1,\dots,n \}$. Here $\hat{u}_i^{\mbox{ NBMQ} \ast}=\hat{u}_h^{\mbox{ NBMQ}}$ and $\hat{\theta}_i^{\mbox{ NBMQ} \ast}=\hat{\theta}_{q_h}^{\mbox{ NBMQ}}$ where $h$ is a random draw from the set $\{1,\dots, n\}$.
\item [step 3.] Generate a bootstrap data set of size $n$, by generating values from a Negative Binomial distribution with
\[
\mu_{i}^\ast=t_i\exp\{\bx_{i}^{T}\hat{\bbeta}_{0.5}+\hat{u}_i^{\mbox{ NBMQ} \ast }\},
\]
\[
\theta_i^\ast=\hat{\theta}_i^{\mbox{ NBMQ} \ast}, ~i=1,\ldots, n,
\]
to obtain bootstrap realizations ${y}_i^\ast$, $i=1, \dots, n$.
\item [step 4.] Fit model (\ref{NBMQModel}) to these bootstrap data and calculate the bootstrap values $\hat{Q}_{q_i}^{\ast} (\bx_i; \psi),~i=1,\dots,n$ of the NBMQ predictors.
\item [step 5.] Repeat steps 2-4 $B$ times. In the $b$-th bootstrap replication, let $y_i^{\ast(b)}$ be the quantity of interest for area $i$, and let $\hat{Q}_{q_i}^{\ast(b)}(\bx_i; \psi)$  be the value of its corresponding NBMQ predictor. The bootstrap estimator of the MSE of (\ref{prediction}) is then
\begin{equation}\label{bootNP}
\mbox{mse}^{\mbox{SPB}}(\hat{Q}_{q_i} (\bx_i; \psi))=B^{-1}\sum_{b=1}^B \Big(\hat{Q}_{q_i}^{\ast(b)}(\bx_i; \psi)- y_i^{\ast(b)} \Big)^2.
\end{equation}
\end{itemize}
Note that this bootstrap procedure can also be used for the NBMQsp predictor by replacing $q_i$ by $q_i^{\mbox{sp}}$.

\section{A model-based simulation study}
\label{sim}

The Scottish lip cancer dataset has been widely analysed in the disease mapping literature (e.g. \cite{clakal}, \cite{brecla} and
\cite{wake}). Here we use these data as the basis for a simulation study that compares the NBMQ and NBMQsp approaches with the EB, BYM and BYMsp approaches to estimating the distribution of relative risk over a defined set of areas.

The data consist of the reported number of lip cancer cases, as well as the
expected number based on population counts, together with an exposure-related covariate indicating the proportion of the population
engaged in agriculture, fishing, or forestry for each of the 56 administrative areas of
Scotland over the period 1975-1980. Following standard practice, we use this proportion divided by ten as the covariate in the model.

The data mechanism used in the simulations emulated the structure of these data, in the sense that independent $y_i\sim \mbox{Poisson}(\mu_i)$ counts were generated based on the model $\mu_i=t_i \exp{(-0.35+0.72x_i+\gamma_i)}$, where the expected cases $t_i$ and covariate values $x_i$ were the same as in the lip cancer dataset, and the values $(-0.35, 0.72)$ used for the model coefficients were defined by the corresponding EB estimates for these data. The heterogeneity effects
$\gamma_i$ were independently drawn from a normal distribution with zero mean and
$\sigma^2$ set equal to $0.15$ or $0.25$. Note that there were no clustering effects in this simulation model, so methods like BYMsp and NBMQsp  that assume the existence of spatial effects can be expected to be relatively inefficient.
In the simulation $1,000$ independent sets of counts were first generated, and each sample was then perturbed by adding $-0.08$ to the
covariate values of four randomly chosen areas (from the $51$ that have a covariate value greater than $0.08$).

Estimated relative risks for the $56$ areas were computed for each set of counts, using the different estimation methods discussed in this paper, i.e. standardised ratios (SMR), Empirical Bayes (EB), Hierarchical Bayes assuming no clustering effects (BYM) and assuming clustering effects (BYMsp), and  Negative Binomial M-quantile modelling, without clustering effects (NBMQ) and allowing for clustering effects (NBMQsp). For each area, the Monte Carlo bias (Bias) and root mean squared error (RMSE) of each estimation procedure was then calculated. The mean values of these performance measures over the $56$ areas are set out in Table \ref{tabella}.
The results  largely confirm our expectations. Under both heterogeneity scenarios ($\sigma^2$=$0.15, 0.25$), EB and BYM report smaller average values of Bias than NBMQ but also higher average values of RMSE, reflecting the usual bias-variance trade-off in outlier-robust estimation. Furthermore, given that the simulated data had no clustering effects, it is not surprising to see that BYMsp is inferior to BYM in terms of average RMSE performance, with virtually identical average Bias. Essentially, there is a variance price to be paid for the overparameterized BYMsp model. However, rather surprisingly, we see that NBMQsp is clearly better than NBMQ in terms of average RMSE, with only a small increase in average Bias. The reason for this becomes clear once one considers the fact that the M-quantile coefficients used in NBMQsp are spatially averaged, see (\ref{sp_qvalue}). This means that the simulated outliers in the data, which have no spatial structure, had much less of an impact on the M-quantile coefficient used by NBMQsp for any particular area. In effect, spatial averaging, in the absence of real clustering in the data, results in M-quantile coefficients that are more stable and shrunk somewhat towards $q=0.5$, leading to lower variability for the corresponding NBMQsp estimates. Of course, this type of shrinkage also implies an increase in average Bias, and this can be seen in the results for NBMQsp in Table \ref{tabella}.

\begin{figure}[!ht]
\hbox{\includegraphics[scale=0.45]{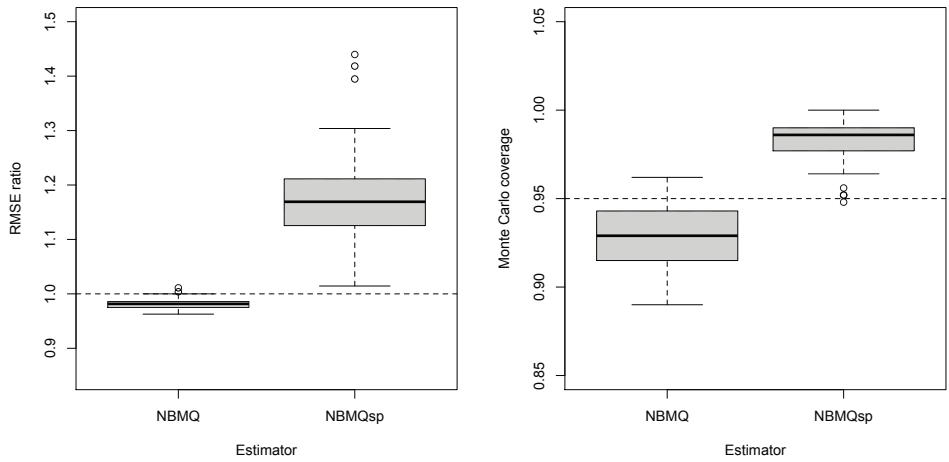}}
\caption{Simulation study: Distribution of MSE ratios (left plot) and Monte Carlo coverages of nominal $95\%$ Gaussian confidence intervals (right plot) generated by semiparametric bootstrap MSE estimator (\ref{bootNP}).}\label{boot}
\end{figure}

We finally examine the performance of the semiparametric bootstrap MSE estimator (\ref{bootNP}). The left hand plot in
Figure \ref{boot} shows the distributions over the 56 areas of the ratios of the Monte Carlo average of (\ref{bootNP}) to the actual Monte Carlo MSE of (\ref{prediction}), while the right hand plot in this Figure shows the distributions of corresponding Monte Carlo coverages of nominal 95\% Gaussian prediction intervals based on (\ref{bootNP}). It is clear that using (\ref{bootNP}) leads to very accurate estimates of the MSE of NBMQ with some undercoverage, while it leads to overestimation of the MSE of NBMQsp with overcoverage.

\section{An application of the NBMQ approach}\label{exe}

We illustrate the NBMQ approach using data on low birth weight incidence
over 2005-2010 for $326$ Local Authority Districts (LADs) in England. The low birth weight data consist
of the number of cases of live and still births with a valid recorded birth-weight of less than 2500 grams.
The data set was obtained
from the UK Public Health Observatory. Expected numbers of cases were defined using internal
standardization based on a set of age-gender specific rates. A preliminary NB-GLM fit of these data indicated
use of the covariates Deprivation Index 2007 and LAD Population Density (defined as population divided by land
area in square miles) for the low birth weight variable.

\begin{figure}
\includegraphics[width=0.75\textwidth]{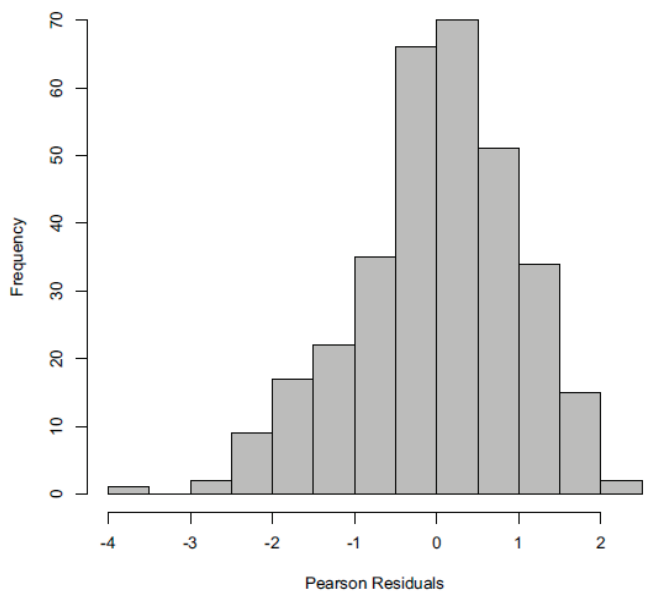}
\caption{Model fit diagnostics for the NB-GLM:
histogram of Pearson residuals for low birth weight data for English LADs.}\label{pearson}
\end{figure}

Figure \ref{pearson} shows the distributions of Pearson residuals generated by NB-GLM fits to the
low birth weight variable using these covariates. These plots indicate the presence of potential
influential observations in the data, with a number of large residuals ($\mid r \mid > 2$)  evident. Further
evidence for the presence of influential observations in these data is obtained when we fit them using robust
NB-GLM and note that although most observations receive a weight of $1$, there are approximately $6.5\%$ that
receive a weight of less than 0.25. We also note that the values of the model covariates are obtained
from UK Public Health Observatory data using small area estimation methods, and so have both sampling
and nonsampling error. Substituting these estimates as covariates in the standard ecological regression models
introduces an additional source of error for LAD-level estimates of low birth weight.
Using an outlier-robust approach, such as one based on an M-quantile model with a bounded influence function,
therefore seems reasonable.

Estimates based on fitting the EB, BYM, BYMsp, NBMQ and NBMQsp ecological regression models to these data were
obtained using \texttt{R} software. The \texttt{eBayes} function in the \texttt{SpatialEpi} library was used
to fit the EB model, while the \texttt{BRugs} library (an \texttt{R} interface to the
\texttt{OpenBUGS} software) was used to fit the BYM and BYMsp models. The NBMQ and NBMQsp models were fitted
using an \texttt{R} function, \texttt{glm.mq.nb} in the \texttt{CountMQ} library that is available from the authors.

\begin{figure}[!ht]
\includegraphics[scale=0.75]{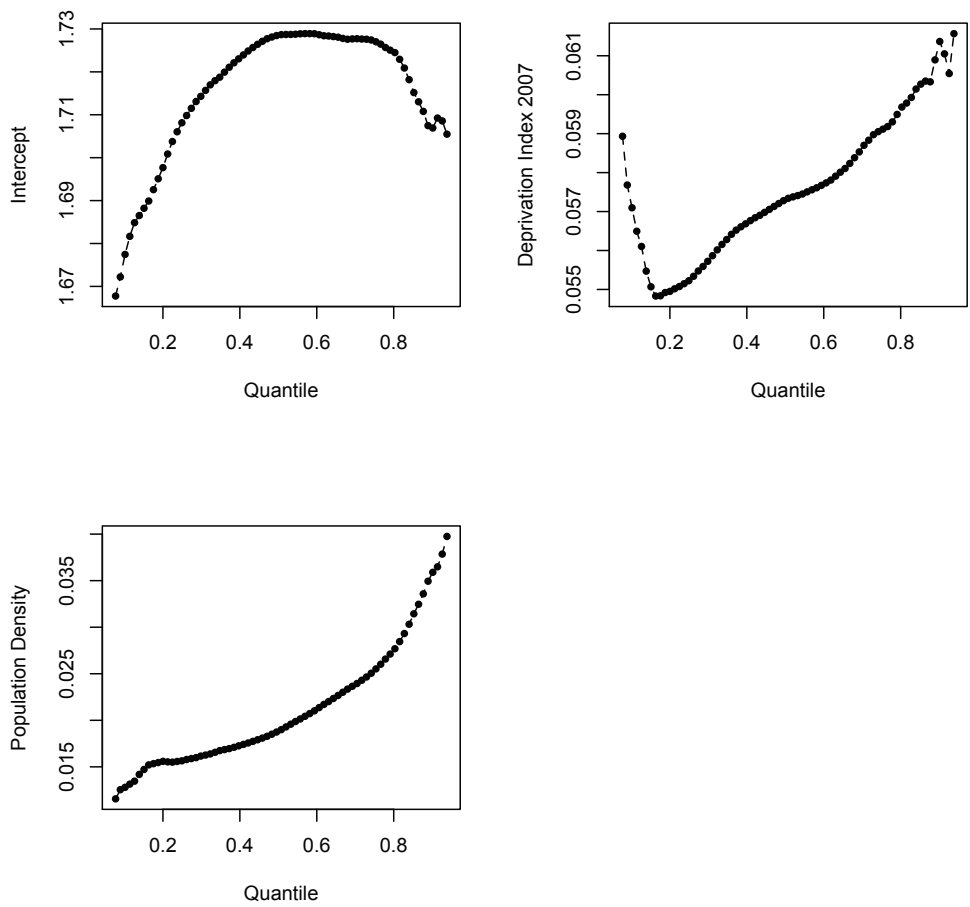}
\caption{Low birth weight data: Regression M-quantile coefficients for the NBMQ model.}\label{betalbw}
\end{figure}

\begin{figure}[!ht]
\includegraphics[scale=0.55]{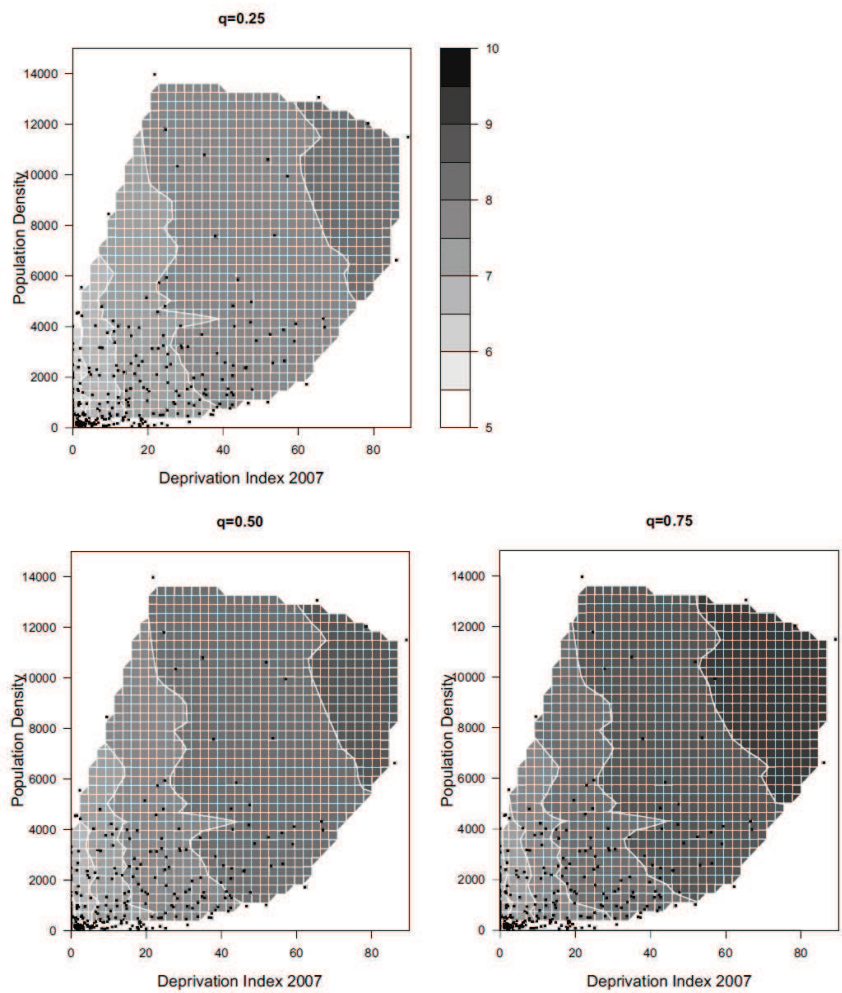}
\caption{Low birth weight data: Contour plots of fitted values generated by the NBMQ model at $q=0.25,~0.50,~0.75$. Individual LADs are shown as points on each contour plot.}\label{contourlwb}
\end{figure}

\begin{figure}[!ht]
\includegraphics[scale=0.75]{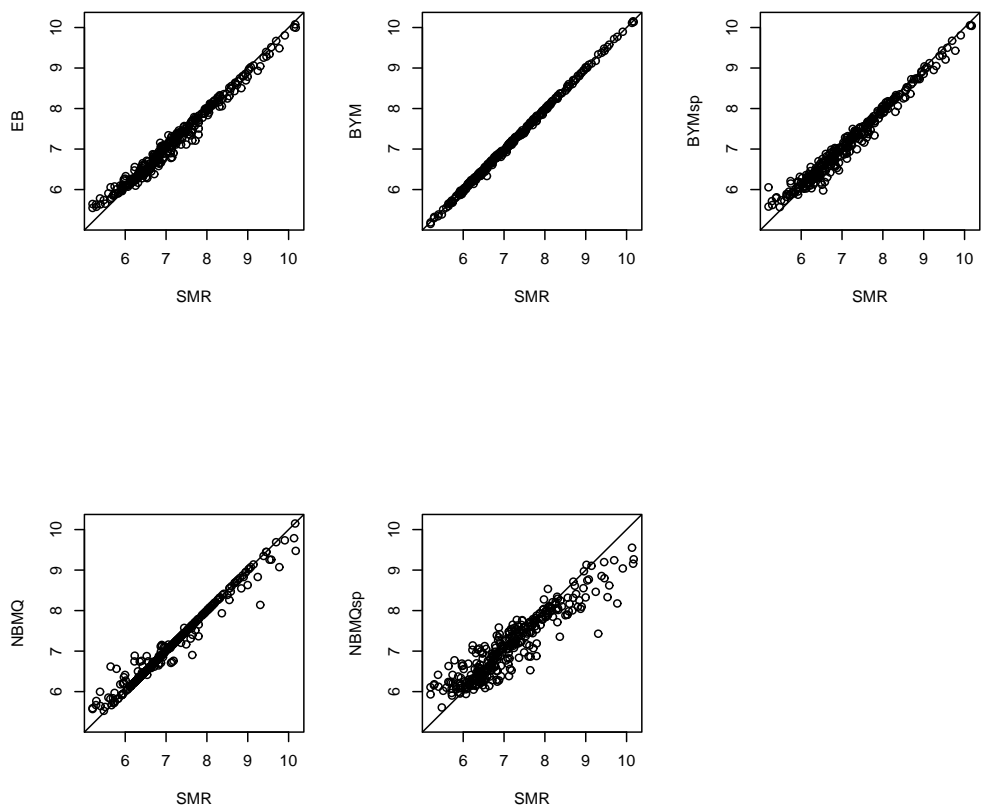}
\caption{Low birth weight data: Estimated relative risks generated by the different approaches plotted against corresponding SMR values.}\label{estimates}
\end{figure}

Figure \ref{betalbw} shows the change in the coefficients of the NBMQ model coefficients
as the quantile index $q$ varies between zero and one.
We see that this change is rather non-linear, particular for values of $q$ near zero and near one,
with the `median' regression M-quantile fit at $q=0.5$ typically quite different from the regression
M-quantile fits at values of $q$ away from this central value. This is evidence of significant overdispersion
in this data set. Furthermore, different covariates have different effects on fitted regression M-quantiles,
as the contour plots in Figure \ref{contourlwb} demonstrate. Here we see that the contours of the fitted
values of the regression M-quantiles of order $q=0.25$, $q=0.5$ and $q=0.75$ all change
faster as Deprivation Index 2007 increases compared with when Population Density increases. Finally,
the scatterplots in Figure \ref{estimates} show the relationships between SMR values and the corresponding
estimates of relative risk generated by the EB, BYM, BYMsp, NBMQ and NBMQsp approaches.
These estimates are in general  agreement, with the smallest correlation (between BYM and  NBMQsp) being $0.93$.
Note, however, that NBMQsp also leads to estimates that appear rather more shrunken towards a common
value than those produced by the other approaches.

\begin{figure}[!h]
\includegraphics[scale=.45]{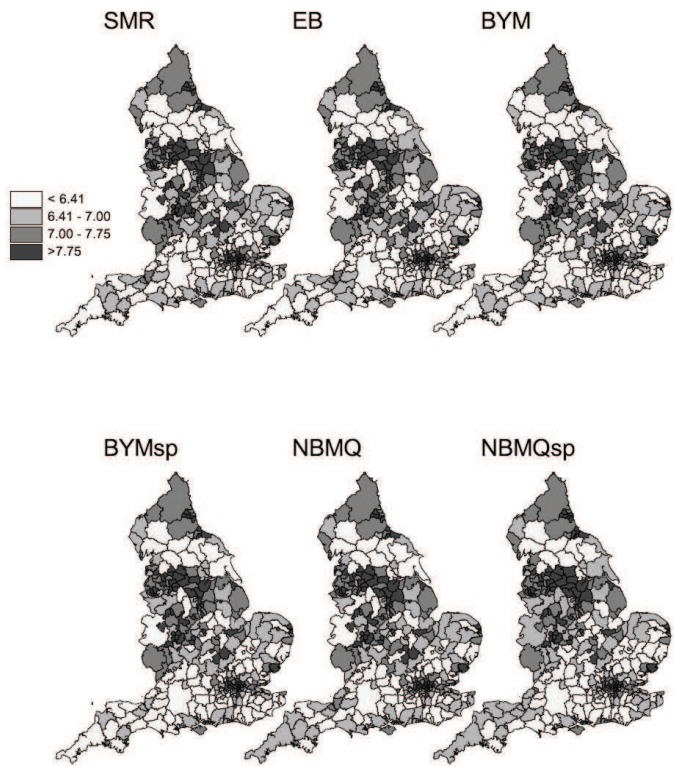}
\caption{Low birth weight data: Maps of estimated relative risks generated by the different approaches.}\label{maplbw}
\end{figure}

\begin{figure}[!h]
\hspace{5cm}\includegraphics[scale=0.3]{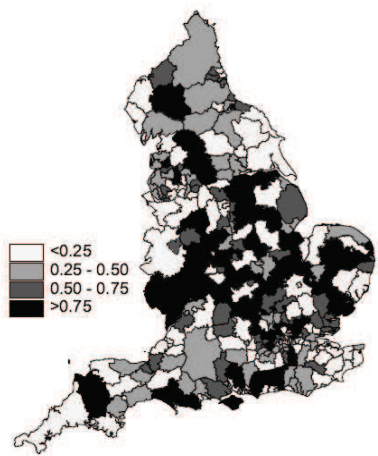}
\caption{Maps of NBMQ M-quantile coefficients $q_i$.}\label{mapq}
\end{figure}

Figure \ref{maplbw} shows the relative risk maps produced by the different approaches.
These are in general agreement, and show the expected geographic
differences due to variation in the model covariates. Risk levels for low birth weight are higher in
urbanized and socio-economic disadvantaged LADs. Of more interest is the spatial distribution of the M-quantile coefficients used in the
NBMQ approach, see equation (\ref{qvalue}), which reflects variability not accounted for by the model
covariates. Figure \ref{mapq} shows this distribution. Here we see pronounced geographic clustering
of these indices, indicating a need for spatial averaging, and hence a preference for
relative risk maps based on NBMQsp.



\section{Conclusion}
\label{con}
We show how an ensemble model defined by the robust regression M-quantiles of a Negative Binomial
distribution can be used to model the count data underpinning disease mapping applications. This modelling approach offers a natural
way of characterising between area variability in count data without imposing
prior assumptions about the source of this variability. In
particular, with an ensemble M-quantile model there is no need to explicitly
specify  the latent variables believed to be the cause of between area variability; rather, inter-area
differences are captured via area-specific M-quantile coefficients.
As a consequence, the need for
distributional assumptions is reduced, and estimation and outlier robust
inference is relatively straightforward. The simulation results that we report in
this paper provide some evidence that the proposed M-quantile modelling approach is a reasonable
alternative to the use of mixed effects models for both ecological analysis and
disease mapping.

However, there remain important issues to be resolved. As with all other model-based methods in current
use for disease mapping, appropriate covariate specification is crucial under the M-quantile modelling
approach, and further research is necessary to develop tools for covariate selection when using ensemble
models like NBMQ and NBMQsp. In this context, we note the work on robust quasi-deviance measures by
Cantoni and Ronchetti \cite{cantoronche}. We also note that the M-quantile modelling approach described here specifically
excludes modelling the quantiles of the count variable of interest, since these are not unique.
Recently, Machado and Santos Silva \cite{machado:2005} and Lee and Neocleous \cite{lee:neo:2010} have proposed an approach to quantile regression
for count data that overcomes this uniqueness problem by jittering the count outcome using additive noise
that is uniformly distributed over the interval $[0,1)$. This form of jittering creates pseudo-smoothness in
the outcome variable and so allows it to be modelled using standard quantile regression methods. Another approach
to quantile modelling of count data was proposed by Efron \cite{efron:1992}, based on the use of asymmetric maximum
likelihood estimation. Further research is necessary to investigate the usefulness of these alternative quantile
regression-based approaches for disease mapping and to compare them with the M-quantile method described here.
Finally, we note that further research is necessary to compare the robustness properties of the NBMQ approach
suggested in this paper with the approach of Bernardinelli \emph{et al.} \cite{berna97}, which explicitly models errors in the covariates.

\section*{Appendix}
\label{appeb}

We evaluate:
\begin{equation*}
\mbox{(i) } E \left[ \psi \left( \frac{
Y_i-\mu_i}{V^{1/2}(\mu_i)}\right)\right]; \, \,  \mbox{(ii) } E \left[ \psi \left( \frac{
Y_i-\mu_i}{V^{1/2}(\mu_i)}\right) \frac{Y_i-\mu_i}{V(\mu_i)}\right]; \, \, \mbox{and (iii) } E \left[ \psi^2 \left( \frac{
Y_i-\mu_i}{V^{1/2}(\mu_i)}\right) \right];
\end{equation*}
where $Y_i$ is distributed according to a NEGBIN2 distribution (see \cite{cametrive}), that is,
\begin{equation*}
P(Y_i=y_i)=\frac{\Gamma(y_i+\theta)}{\Gamma(\theta)\,y_i!}\,\left(\frac{\mu_i}{\mu_i+\theta}\right)^{y_i}\,\left(\frac{\theta}{\mu_i+\theta}\right)^\theta \, \, \mbox{for} \, \, y_i=0,1,2,\ldots
\end{equation*}
Here $\theta$ is a positive integer, $\mu_i=E(Y_i)$ and $V(\mu_i)=\mbox{var}(Y_i)=\mu_i+\frac{\mu_i^2}{\theta}$. To simplify the notation, the index $i$ is suppressed from now on.

First, we evaluate $E\bigl[\,Y\,I(Y\in A)\bigr]$ and $E\bigl[\,Y^2\,I(Y\in A)\bigr]$, where $A=\{a,\ldots,b-1\}$ and $0 \leq a < b$ are integers.
Let $A+1=\{a+1,\ldots,b\}$.
Then
\[
E\bigl[\,Y\,I(Y\in A+1)\bigr]=E\bigl[Y\,I(Y\in A)\bigr]-a\,P(Y=a)+b\,P(Y=b).
\]
Using the transformation $z=y-1$, one also obtains
\[
\begin{array}{rl}
  E\bigl[\,Y\,I(Y\in A+1)\bigr] & =\sum_{y\in A+1}y\,\frac{\Gamma(y+\theta)}{\Gamma(\theta)\,y!}\,\bigl(\frac{\mu}{\mu+\theta}\bigr)^y\,\bigl(\frac{\theta}{\mu+\theta}\bigr)^\theta\\
   & =\frac{\mu}{\mu+\theta}\,\sum_{z\in A}(z+\theta)\,\frac{\Gamma(z+\theta)}{\Gamma(\theta)\,z!}\,\bigl(\frac{\mu}{\mu+\theta}\bigr)^z\,\bigl(\frac{\theta}{\mu+\theta}\bigr)^\theta\\
   & =\frac{\mu}{\mu+\theta}\,\sum_{z\in A}(z+\theta)\,P(Y=z) \\
   & =\frac{\mu}{\mu+\theta}\,E\bigl[\,Y\,I(Y\in A)\bigr]+\frac{\mu\,\theta}{\mu+\theta}\,P(Y\in A).
\end{array}
\]
Equating these expressions, we see that
\begin{equation}\label{atteso1}
E\left[Y\,I(Y\in A)\right]=\frac{\mu+\theta}{\theta}\,\left[a P(Y=a)- b P(Y=b)\right]+\mu \,P(Y\in A).
\end{equation}
We next apply the same argument to evaluating $E\left[Y^2\,I(Y\in A)\right]$. In this case,
\begin{equation*}
E\left[\,Y^2\,I(Y\in A+1)\right]=E\left[Y^2\,I(Y\in A)\right]-a^2\,P(Y=a)+b^2\,P(Y=b)
\end{equation*}
and
\[
\begin{array}{rl}
E\left[Y^2\,I(Y \in A+1)\right]&=
\sum_{y\in A+1}y^2\,\frac{\Gamma(y+\theta)}{\Gamma(\theta)\,y!}\,\bigl(\frac{\mu}{\mu+\theta}\bigr)^y\,\bigl(\frac{\theta}{\mu+\theta}\bigr)^\theta\\
&=\frac{\mu}{\mu+\theta}\,\sum_{z\in A}(z+1)(z+\theta)\,\frac{\Gamma(z+\theta)}{\Gamma(\theta)\,z!}\,\bigl(\frac{\mu}{\mu+\theta}\bigr)^z\,\bigl(\frac{\theta}{\mu+\theta}\bigr)^\theta\\
&=\frac{\mu}{\mu+\theta}\,\sum_{z\in A}\bigl(z^2+(\theta+1)z+\theta\bigr)\,P(Y=z)\\
&=\frac{\mu}{\mu+\theta}\,E\bigl[\,Y^2\,I(Y\in A)\bigr]+\frac{\mu\,(\theta+1)}{\mu+\theta}\,E\bigl[\,Y\,I(Y\in A)\bigr]+\frac{\mu\,\theta}{\mu+\theta}\,P(Y\in A).
\end{array}
\]
Again, equating these expressions yields
\begin{equation}\label{atteso2}
\begin{array}{rl}
E\left[Y^2 \, I(Y \in A)\right]&=\frac{\mu+\theta}{\theta}\,\left[a^2 P(Y=a)- b^2 P(Y=b)\right]+
\frac{\mu (\theta+1)}{\theta}\,E \left[Y\,I(Y\in A)\right]+\mu\,P(Y\in A)=\\
&=\frac{\mu}{\theta}\,\left[\theta+\mu \theta+\mu\right] P(Y\in A)+\frac{\mu+\theta}{\theta}\,\left[a^2\,P(Y=a)-b^2\,P(Y=b)\right]+\\
&+\frac{\mu (\mu+\theta) (\theta+1)}{\theta^2}\left[a P(Y=a)- b P(Y=b)\right].\\
\end{array}
\end{equation}

We are now in a position to evaluate (i)$-$(iii) given $A$. Define
\[
\psi(r)=\left\{\begin{array}{ll}
                     r   & -c \leq r \leq c \\
                     c   & r > c \\
                     -c  & r < -c
                   \end{array} \right.
\]
and set $r=\frac{Y-\mu}{V^{1/2}(\mu)}$. Let $j_1=\lfloor \mu - c \, V^{1/2}(\mu)\rfloor$ and $j_2=\lfloor \mu+ c \, V^{1/2}(\mu)\rfloor$.
Note that the results obtained below may change depending on whether or not $\mu - c \, V^{1/2}(\mu)$ is an integer.
In what follows, we assume the non-integer case; the integer case can be handled similarly.

\begin{itemize}
\item[(i)]
\[
E \left[ \psi \left( \frac{Y-\mu }{V^{1/2}(\mu )}\right)\right]= -
c \, P\left( \frac{ Y -\mu }{V^{1/2}(\mu )} < -c \right) + c \,
P\left(\frac{ Y -\mu }{V^{1/2}(\mu )} > c\right) +
\]
\[
+E \left[\frac{ Y -\mu }{V^{1/2}(\mu )} \, \, I(-c \leq \frac{ Y -\mu }{V^{1/2}(\mu )} \leq c)\right]=
\]
Since $\frac{Y-\mu}{V^{1/2}(\mu)} > c$ implies $Y  > \mu+ c \,V^{1/2}(\mu)$, as $Y$ is integer valued, we have
$Y \geq \lfloor \mu+c \,V^{1/2}(\mu)\rfloor +1=j_2+1$.
Analogously, $\frac{ Y -\mu }{V^{1/2}(\mu )} < -c$ implies $Y < \mu - c \,V^{1/2}(\mu )$, which, since $\mu - c \,V^{1/2}(\mu )$
is not integer, leads to $Y \leq \lfloor \mu -
c \,V^{1/2}(\mu )\rfloor =j_1$ (when $\mu - c \,V^{1/2}(\mu )$ is integer to $Y \leq j_1-1$).
Moreover, $-c \leq \frac{ Y -\mu }{V^{1/2}(\mu )} \leq c$ implies  $\mu - c \, V^{1/2}(\mu ) \leq Y \leq \mu + c \, V^{1/2}(\mu )$
which amounts to $j_1+1 \leq Y \leq j_2$ (when $\mu - c \,V^{1/2}(\mu )$ is integer is $j_1 \leq Y \leq j_2$). So
\[
E \left[ \psi \left( \frac{Y-\mu }{V^{1/2}(\mu )}\right)\right]=-c \, P\left(Y \leq j_1\right) + c \, P \left(Y \geq
j_2+1 \right) +
\]
\[
+ \frac{1}{V^{1/2}(\mu )} E \left[ Y \, I(j_1+1 \leq Y
\leq j_2)\right] - \frac{\mu}{V^{1/2}(\mu)} P(j_1+1 \leq Y
\leq j_2).
\]
Considering $A=\{j_1+1, \ldots,j_2\}$ and also that
\begin{equation}\label{passo}
\frac{\mu +\theta}{\theta} (y+1) P(Y=y+1)=\frac{\mu}{\theta} y
P(Y=y)+\mu P(Y=y)
\end{equation}
we obtain
\begin{equation}\label{atteso1p}
E \left[ Y \, I(j_1+1 \leq Y \leq j_2)\right]= \mu P( j_1 \leq Y \leq j_2-1)- \frac{\mu }{\theta} j_2
P(Y =j_2)+\frac{\mu }{\theta} j_1 P(Y =j_1)
\end{equation}
and finally
\[
\begin{array}{rl}
E \left[ \psi \left( \frac{Y -\mu }{V^{1/2}(\mu )}\right)\right]&=-c \, P\left(Y \leq j_1\right) + c \, P \left(Y \geq j_2+1 \right)+\\
&+\frac{\mu }{V^{1/2}(\mu )} P( Y =j_1) \left( 1+\frac{j_1}{\theta}\right)
- \frac{\mu }{V^{1/2}(\mu )} P(Y =j_2) \left(1+\frac{j_2}{\theta} \right).\\
\end{array}
\]

\item[(ii)]
\[
E \left[ \psi \left( \frac{
Y -\mu }{V^{1/2}(\mu )}\right) \frac{Y -\mu }{V(\mu )}\right]=
-\frac{c}{V(\mu )} E \left[(Y - \mu ) \, I(Y \leq j_1)\right]+\frac{c}{V(\mu )} E \left[(Y - \mu ) \, I(Y \geq j_2+1)\right]+
\]
\[
+\frac{1}{V^{3/2}(\mu )} E \left[(Y - \mu )^2 \, I(j_1+1 \leq Y \leq j_2)\right]=
\]
\[
=\frac{\mu \, c}{V(\mu )} P(Y \leq j_1)+\frac{\mu \, c}{V(\mu )} P(Y \leq j_2)
+\frac{\mu ^2}{V^{3/2}(\mu )} P \left(j_1+1 \leq Y \leq j_2 \right)-
\]
\[
-\frac{c}{V(\mu )} E\left[ Y \, I(Y \leq j_1)\right]-\frac{c}{V(\mu )} E\left[ Y \, I(Y \leq j_2)\right]
-\frac{2 \, \mu }{V^{3/2}(\mu )} E \left[Y \,  I(j_1+1 \leq Y \leq j_2)\right]+
\]
\[
+\frac{1}{V^{3/2}(\mu )} E \left[Y ^2 \,  I(j_1+1 \leq Y \leq j_2)\right]
\]
From (\ref{atteso1p}), setting $A=\{0, \ldots, j_1\}$ in (\ref{atteso1}) and using (\ref{passo}) leads to
\begin{equation}\label{e2}
E \left[Y \, I(Y \leq j_1) \right]=-\frac{\mu }{\theta} j_1 \, P(Y =j_1)+ \mu \, P(Y \leq j_1-1).
\end{equation}
Similarly, setting $A=\{0, \ldots, j_2\}$ in (\ref{atteso1}) and using (\ref{passo}) gives
\begin{equation}\label{e3}
E \left[ Y \, I(Y \leq j_2) \right]=-\frac{\mu }{\theta} j_2 \, P(Y =j_2)+ \mu \, P(Y \leq j_2-1).
\end{equation}
Finally, setting $A=\{j_1+1, \ldots, j_2\}$ in (\ref{atteso2}) and using
(\ref{passo}) leads to
\begin{equation}\label{atteso2p}
E \left[Y^2 \, I(j_1+1 \leq Y \leq j_2)\right]=
\frac{\mu}{\theta} \, (\theta+\mu \, \theta+\mu) \, P(j_1+1 \leq Y \leq j_2)
\end{equation}
\[
+\left[j_1+1+\frac{\mu (\theta+1)}{\theta}\right] \left[\frac{\mu}{\theta}\,  j_1 \, P(Y=j_1)+\mu \, P(Y=j_1)\right]
\]
\[
-\left[j_2+1+\frac{\mu (\theta+1)}{\theta}\right] \left[\frac{\mu}{\theta} \, j_2 \, P(Y=j_2)+\mu \, P(Y=j_2)\right].
\]
Finally, substituting (\ref{atteso1p}), (\ref{e2}), (\ref{e3}) and (\ref{atteso2p}),
\[
E \left[ \psi \left( \frac{Y -\mu }{V^{1/2}(\mu )}\right) \frac{Y
-\mu }{V(\mu )}\right]= \frac{\mu \, c}{V(\mu )} \left[ P(Y =j_1)
\frac{j_1+\theta}{\theta}+ P(Y =j_2)
\frac{j_2+\theta}{\theta}\right]
\]
\[
+\frac{\mu }{V^{3/2}(\mu )} \left[ P(Y =j_1)  \frac{j_1}{\theta} \,
(\theta+1+j_1) - P(Y =j_2) \frac{j_2}{\theta} \, (\theta+1+j_2) +
P(j_1 \leq Y \leq j_2-1) \right]
\]
\[
+\frac{\mu ^2}{V^{3/2}(\mu )} \left\{P(Y =j_1) \left[\frac{j_1-j_1
\theta - \theta^2}{\theta^2}\right] -
 P(Y =j_2) \left[ \frac{j_2-j_2 \theta - \theta^2}{\theta^2}\right] + \frac{1}{\theta} P(j_1 \leq Y \leq j_2-1)\right\}.
\]

\item[(iii)]
\[
E \left[ \psi^2 \left( \frac{
Y -\mu }{V^{1/2}(\mu )}\right) \right]=c^2 \left[ P(Y \leq j_1) + P(Y \geq j_2+1) \right]+
\]
\[
+\frac{1}{V(\mu )} E \left[ (Y-\mu )^2 \, I(j_1+1 \leq Y
\leq j_2)\right]=
\]
\[
=c^2 \left[1-P(j_1+1 \leq Y \leq j_2) \right]+ \frac{\mu ^2}{V(\mu )} P(j_1+1 \leq Y \leq j_2)
\]
\[
- \frac{2 \mu }{V(\mu )} E[Y \, I(j_1+1 \leq Y \leq j_2 )]+ \frac{1}{V(\mu )} E[Y ^2 \, I(j_1+1 \leq Y \leq j_2)].
\]
Substituting the expected values (\ref{atteso1p}) and ( \ref{atteso2p}) we obtain
\[
E \left[ \psi^2 \left( \frac{ Y -\mu }{V^{1/2}(\mu )}\right)
\right] = c^2 \, \, [1- P(j_1+1 \leq Y \leq j_2)]+
\]
\[
+\frac{\mu }{V(\mu )} \left[ P(Y =j_1) \, \,  \frac{j_1}{\theta} \,
(\theta+1+j_1) - P(Y =j_2) \, \,  \frac{j_2}{\theta} \,
(\theta+1+j_2) + P(j_1 \leq Y \leq j_2-1) \right]+
\]
\[
+\frac{\mu ^2}{V(\mu )} \left\{ P(Y =j_1) \left[ \frac{j_1-j_1
\theta - \theta^2}{\theta^2}\right] -
 P(Y =j_2) \left[ \frac{j_2-j_2 \theta - \theta^2}{\theta^2}\right] + \frac{1}{\theta}  P(j_1 \leq Y \leq j_2-1) \right\}.
\]
\end{itemize}

\textbf{Acknowledgement:} The authors are grateful to William Aeberhard for helpful comments.

\newpage

\begin{table}
\center
\caption{Model-based simulation results: Mean values of Monte Carlo biases and RMSEs of
predictors of relative risk.}\label{tabella}
\begin{tabular}{ccccc}
      & \multicolumn{2}{c}{ $\sigma^2$=$0.15$} &\multicolumn{2}{c}{ $\sigma^2$=$0.25$}\\
      &  Bias    & RMSE  & Bias   & RMSE\\
      \hline
EB    &  -0.006   & 0.520 & -0.013 & 0.759\\
BYM   &  -0.004   & 0.539 & -0.012 & 0.784\\
BYMsp &  -0.003   & 0.560 & -0.013 & 0.800\\
NBMQ  &  -0.030   & 0.398 & -0.061 & 0.499\\
NBMQsp&  -0.032   & 0.280 & -0.063 & 0.352 \\\hline
\end{tabular}
\end{table}

\newpage

\newpage

\newpage

\newpage

\newpage

\newpage

\end{document}